\documentclass[journal]{IEEEtran}
\usepackage{cite}
\ifCLASSINFOpdf
   \usepackage[pdftex]{graphicx}
   \graphicspath{{../pdf/}{../jpeg/}}
   \DeclareGraphicsExtensions{.pdf,.jpeg,.png}
\else
  \usepackage[dvips]{graphicx}
  \graphicspath{{../eps/}}
  \DeclareGraphicsExtensions{.eps}
\fi

\usepackage[cmex10]{amsmath}
\usepackage[caption=false,font=footnotesize]{subfig}
\usepackage{fixltx2e}
\usepackage{dblfloatfix}
\usepackage{epstopdf}

\begin{document}
%
\title{On the Parameter Estimation of Sinusoidal Models for Speech and Audio Signals}
%
%
%

\author{George~P.~Kafentzis,~\IEEEmembership{Member,~IEEE}}

%
%

\markboth{Journal of \LaTeX\ Class Files,~Vol.~11, No.~4, December~2023}%
{Shell \MakeLowercase{\textit{et al.}}: Bare Demo of IEEEtran.cls for Journals}
%



\maketitle

\begin{abstract}
In this paper, we examine the parameter estimation performance of three well-known sinusoidal models for speech and audio. The first one is the standard Sinusoidal Model (SM), which is based on the Fast Fourier Transform (FFT). The second is the Exponentially Damped Sinusoidal Model (EDSM) which has been proposed in the last decade, and utilizes a subspace method for parameter estimation, and finally the extended adaptive Quasi-Harmonic Model (eaQHM), which has been recently proposed for AM-FM decomposition, and estimates the signal parameters using Least Squares on a set of basis function that are adaptive to the local characteristics of the signal. The parameter estimation of each model is briefly described and its performance is compared to the others in terms of signal reconstruction accuracy versus window size on a variety of synthetic signals and versus the number of sinusoids on real signals. The latter include highly non stationary signals, such as singing voices and guitar solos. The advantages and disadvantages of each model are presented via synthetic signals and then the application on real signals is discussed. Conclusively, eaQHM outperforms EDS in medium-to-large window size analysis, whereas EDSM yields higher reconstruction values for smaller analysis window sizes. Thus, a future research direction appears to be the merge of adaptivity of the eaQHM and parameter estimation robustness of the EDSM in a new paradigm for high-quality analysis and resynthesis of general audio signals.
\end{abstract}


%
\IEEEpeerreviewmaketitle

\section{Introduction}
\label{sec:intro}
Sinusoidal modeling is a renowned parametric representation for speech and audio signals. Sinusoidal models have been used in the fields of speech and audio signal processing in coding~\cite{George:87, mcaulay-92, Ahmadi:01}, analysis and synthesis~\cite{McAulay:86, Serra:89, Quatieri:02, Laroche:93, George:97, Pantazis:10, Caetano:13a}, enhancement~\cite{deisher:97a, Jensen:01, Zavarehei:07, Stark:08}, modifications and transformations~\cite{Serra:89, George:97, Quatieri:92, Stylianou:95, Osaka:95, Federico:98, Kafentzis:13}, among others. 

The general problem of fitting a sum of sinusoids to a signal is of great interest. Thus, many algorithms have been developed for accurate estimation of the sinusoidal parameters. When it comes to speech and audio, the algorithms can be separated into four categories: spectral analysis~\cite{McAulay:86, Serra:89}, analysis-by-synthesis~\cite{Goodwin:97, George:97, Verma:99, Heusdens:02}, least-squares minimization~\cite{Stylianou:phd:96, Pantazis:11, Degottex:13}, and subspace methods~\cite{Nieuwenhuisje:98, Jensen:04, Badeau:06}.

In general, a sinusoidal model (SM) can be described as
\begin{equation}
s(t) = \sum_{k=1}^{K}A_k(t)\cos(\phi_k(t))
\label{eq:SM}
\end{equation}
where $A_k(t)$ and $\phi_k(t)$ are the instantaneous amplitude and phase of the $k^{th}$ sinusoid. First approaches on parameter estimation for sinusoidal modeling relied on the assumption of local stationarity, that is, signals were considered as stationary in amplitude and frequency in short time intervals, i.e. $20-30$ ms~\cite{McAulay:86, Serra:89}.
The estimation of these components are performed under the aforementioned assumption of stationarity using standard spectral analysis with the Fast Fourier Transform (FFT). These approaches suffered from the limited time-frequency resolution of the FFT. Nevertheless, the model has been successfully applied in audio and speech signal analysis, but is far from being valid in signals which are highly non-stationary, such as speech onsets and sharp attacks in music signals, and consequently performs rather poorly.

Since then, better models and parameter estimation methods have been developed to account for the basic problems of parameter estimation, and are considered as generalizations of the basic sinusoidal model. One of them is the so-called Exponentially Damped Sinusoidal Model (EDSM)~\cite{Nieuwenhuisje:98, Jensen:99}. This model allows for the amplitude of each sinusoid to vary exponentially with time. Such a model can be described as
\begin{equation}
s(t) = \sum_{k=1}^K a_ke^{-d_k t} \cos(\omega_k t + \phi_k)
\label{eq:EDSM}
\end{equation}
where $t \in [0,L-1]$, and the $d_k$ term represents the damping of the sinusoids. Please note that when $d_k=0$, then EDSM reduces to the standard Sinusoidal Model of Eq.~(\ref{eq:SM}). Parameter estimation in EDSM is performed using Matching Pursuit or subspace methods which have good spectral properties and do not suffer from the time-frequency trade-off embedded in the aforementioned methods. However, EDS still considers frequency stationarity inside the analysis window.

More recently, adaptive Sinusoidal Models (aSMs)~\cite{Pantazis:11, Kafentzis:12, Degottex:13} have gained attention due to their ability to adapt their parameters to the local characteristics of the signal, thus alleviating the stationarity problem via an iterative parameter re-estimation process, the so-called adaptation. An aSM, although it can be harmonic~\cite{Degottex:13} or quasi-harmonic~\cite{Pantazis:11, Kafentzis:12}, can be generally described similarly to the SM, as a sum of time-varying AM-FM components (Eq.~(\ref{eq:SM})). The parameters are estimated using Least-Squares minimization and the adaptation process enhances their accuracy. The aSMs have shown to perform well on studio recordings of speech signals~\cite{Pantazis:10, Degottex:13} and on single tones of musical instrument sounds~\cite{Caetano:13a, Caetano:13b}. However, there is no investigation on their performance on highly non-stationary running audio samples, such as singing voice, guitar solos, pitch-varying musical instruments, etc.

In this paper, we examine the standard SM, with the FFT as a parameter estimator, a subspace-based EDSM, and an adaptive Sinusoidal Model called the extended adaptive Quasi-Harmonic Model (eaQHM) on synthetic signals first, in order to reveal insightful information on the performance characteristics and the advantages of each model. Then, the models analyze real signals, and particularly monophonic, running audio and speech signals based on the observations made on synthetic signals.

The paper is organized as follows. Section $2$ briefly describes the EDS Model and a subspace method for its parameter estimation and Section $3$ reviews the theory of the extended adaptive Quasi-Harmonic Model. Section $4$ presents the evaluation on synthetic and real signals and Section $5$ discusses the results. Finally, Section $6$ concludes the paper.

\section{Exponentially Damped Sinusoidal Model (EDSM)}
The EDS model for a signal $x(t)$, with support in $[0,L]$, can be written as
\begin{equation}
x(t) = \sum_{k=0}^{K-1}s_k(t)
\label{eq:EDS}
\end{equation}
where $L$ is the length of the signal segment and $K$ is the model order (number of sinusoids), $s_k(t)$ is the $k^{th}$ exponentially damped sinusoid. The exponentially damped sinusoid $s_k(t)$ can be written as
\begin{equation}
s_k(t) = \alpha_k z_k^t
\end{equation}
where $\alpha_k$ are the complex amplitudes, and $z_k$ are the complex poles. Furthermore, the poles can be written as
\begin{equation}
z_k = e^{\frac{1}{L}(\delta_k + i\omega_k)}
\end{equation}
where $\delta_k$ are the damping factors and $\omega_k$ are the angular frequencies (or pulsations). The damping factor is directly related to the sharpness of the exponential time envelope, where a positive damping factor corresponds to an increasing envelope, whereas a negative damping factor reflects a decreasing envelope. A zero damping factor corresponds to an amplitude-stationary sinusoid. Also, the amplitude can be written as
\begin{equation}
\alpha_k = \left\{\begin{array}{ll}
a_ke^{-\delta_k + i\phi_k}, & \delta_k \geq 0\\
a_ke^{i\phi_k}, & \delta_k < 0 
\end{array}\right.
\end{equation}
where $a_k$ are the real amplitudes and $\phi_k$ the phases. The different expressions of $\alpha_k$ helps numerical stability.
Compared to the most common methods for parameter estimation in audio and speech signal analysis, such as the FFT and the Least-Squares, the EDS uses subspace methods to efficiently estimate its parameters~\cite{Nieuwenhuisje:98, Jensen:04, Hermus:05, Badeau:06}. Among different implementations of EDS estimation schemes, in this paper we use the extension of ESPRIT described in~\cite{Badeau:06}. A description of the algorithm is briefly presented next.

\subsection{Parameter Estimation}
Let us define the signal vector
\begin{equation}
\mathbf{x} = [ x[0] \quad x[1] \quad \cdots \quad x[L-1] ]^T
\end{equation}
where $(\cdot )^T$ denotes transposition. Then, the Hankel signal matrix is defined as
\begin{equation}
\mathbf{X} = \begin{bmatrix}
x[0] & x[1] & \cdots & x[N-1] \\
x[1] & x[2] & \cdots & x[N] \\
\vdots & \vdots & \vdots & \vdots\\
x[R-1] & x[R] & \cdots & x[L-1]
\end{bmatrix}
\end{equation}
where $N > K$, $R > K$ and $N + R - 1 = L$. A value of $N = L/2$ is suggested as efficient in~\cite{20}. Moreover, the complex-amplitude vector is defined
\begin{equation}
\mathbf{\alpha} = [ \alpha_0 \quad \alpha_1 \quad \cdots \quad \alpha_{K-1} ]^T
\end{equation}
and the so-called Vandermode matrix of the poles as
\begin{equation}
\mathbf{Z^T} = \begin{bmatrix}
1 & 1 & \cdots & 1 \\
z_0 & z_1 & \cdots & z_{K-1} \\
\vdots & \vdots & \vdots & \vdots\\
z_0^{L-1} & z_1^{L-1} & \cdots & z_{K-1}^{L-1}
\end{bmatrix}
\end{equation}
Using the above expressions, Eq.~\ref{eq:EDS} can be written as
\begin{equation}
\mathbf{x} = \mathbf{Z^T}\mathbf{\alpha}
\label{eq:EDSsynth}
\end{equation}
The objective is to first find the poles ad then the optimal amplitudes that correspond to those poles. Doing so, a Singular Value Decomposition (SVD) on matrix $\mathbf{X}$ gives
\begin{equation}
\mathbf{X} = [\mathbf{U_1} \: \: \mathbf{U_2}]\begin{bmatrix}
\mathbf{\Sigma_1} & \mathbf{0} \\
\mathbf{0} & \mathbf{\Sigma_2}
\end{bmatrix}\begin{bmatrix}
\mathbf{V_1} \\
\mathbf{V_2}
\end{bmatrix}
\end{equation}
where $\mathbf{\Sigma_1}$ and $\mathbf{\Sigma_2}$ are diagonal matrices containing the $K$ largest singular values and the smallest singular values, respectively. Matrices $[\mathbf{U_1 \: U_2}]$ and $[\mathbf{V_1 \: V_2}]$ are the corresponding left and right singular vector matrices, respectively. Moreover, the shift-invariance property of the signal space spanned by $V_1$ gives
\begin{equation}
\mathbf{V_1^{\downarrow}\Phi} = \mathbf{V_1^{\uparrow}}
\end{equation}
Notation $(\cdot )^{\uparrow \downarrow}$ stands for the operations of removing the first and last line of a matrix, respectively. Matrix $\mathbf{\Phi}$ has eigenvalues that correspond to the poles, $z_k$. A mean-squared error is set for the calculation of matrix $\mathbf{\Phi}$, and thus $\mathbf{\Phi}$ is obtained by
\begin{equation}
\mathbf{\Phi} = \Big(\mathbf{V_1^{\downarrow}}\Big)^+\mathbf{V_1^{\uparrow}}
\end{equation}
where $(\cdot)^+$ denotes the pseudo-inverse of a matrix. Then, the poles $z_k$ are obtained from a diagonalization of $\mathbf{\Phi}$. Having a set of poles, the Vandermonde matrix $\mathbf{Z^T}$ can be computed, and thus the optimal amplitudes are obtained by
\begin{equation}
\mathbf{\alpha} = \Big(\mathbf{Z^T}\Big)^+\mathbf{x}
\end{equation}
Finally, each frame can be synthesized on a frame-based manner using Eq.~(\ref{eq:EDSsynth}).

\section{Introduction to the Adaptive Sinusoidal Models - aSMs}
Contrary to subspace or FFT-based methods, the aSMs utilize the Least-Squares minimization criterion to estimate the parameters. The \textit{adaptive} term is justified by a sequence of parameter re-estimation based on successive refinements of the model basis functions. 

In general, an aSM can be described as
\begin{equation}
\begin{aligned}
x(t) = \left( \sum_{k=-K}^{K}C_k(t)\psi_k(t) \right)w(t)
\label{eq:ASM:eaQHMeq}
\end{aligned}
\end{equation}
where $\psi_k(t)$ denotes the set of basis functions, $C_k(t)$ denotes the (complex) amplitude term of the model, $2K+1$ is the number of exponentials (hence, $K+1$ sinusoids), and finally $w(t)$ is the analysis window with support in $[-T,T]$. 

Using this notation, in conventional sinusoidal models including the SM, the EDSM, the Harmonic Model (HM)~\cite{Stylianou:phd:96}, the Quasi-Harmonic Model (QHM)~\cite{Pantazis:08c}, and others, the set of basis functions $\psi_k(t)$ in the analysis part is stationary in frequency and in amplitude. For example, the basis functions in the SM are in the form of
\begin{equation}
\psi_k^{SM}(t) = 1 \cdot e^{j2\pi f_k t}, \: \: C_k^{SM}(t) = a_k
\end{equation}
where the amplitudes and frequencies of the basis functions are constant inside the analysis window ($1$ and $f_k$, respectively). Similarly, the basis functions in the HM are
\begin{equation}
\psi_k^{HM}(t) = 1 \cdot e^{j2\pi kf_0 t}, \: \: C_k^{HM}(t) = a_k
\end{equation}
where the frequencies are restricted to be harmonically related. 
It is clear that in both models the basis functions $\psi_k(t)$ consist of exponential functions that are \textit{stationary} in amplitude and frequency, due to the inherent assumption that the signal does not significantly change in short time intervals. On the contrary, the aSMs do not share this assumption, as will be explained in the next section.

\subsection{The extended adaptive Quasi-Harmonic Model - \lowercase{ea}QHM}
In this paragraph, a recently developed adaptive Sinusoidal Model, named the extended adaptive Quasi-Harmonic Model (eaQHM), will be briefly described.

The eaQHM projects a signal segment onto a set of non-parametric, time-varying basis functions with instantaneous amplitudes and phases that are adaptive to the local characteristics of the underlying signal~\cite{Kafentzis:12}. Specifically, in eaQHM,
\begin{equation}
\psi_k^{eaQHM}(t) = \hat{\alpha}_k(t)e^{j\hat{\phi}_k(t)}, \: \: C_k^{eaQHM}(t) = (a_k + tb_k)
\end{equation}
where $a_k$ and $b_k$ are the complex amplitude and the complex slope of the model respectively, $\hat{\alpha}_k(t)$ and $\hat{\phi}_k(t)$ are estimates of the instantaneous amplitude and phase of the $k^{th}$ component, respectively, both obtained from an initialization step.

The instantaneous phase is computed using a frequency integration scheme~\cite{Pantazis:11}, although cubic phase interpolation could be used as well~\cite{McAulay:86}. The eaQHM is actually a parameter-refinement mechanism, thus it requires an initialization, as already mentioned. For this purpose, any AM-FM decomposition algorithm can be used, but in most of the previous works concerning the eaQHM~\cite{Kafentzis:aSM:14, Pantazis:11}, the Harmonic Model (HM)~\cite{Stylianou:phd:96} or the Quasi-Harmonic Model is used.

\subsection{Parameter Estimation}
\label{sec:eaQHM}
Considering that a preliminary estimation of the instantaneous components $\alpha_k(t)$ and $\phi_k(t)$ of the signal is available, the estimation of the unknown parameters of eaQHM is similar to that of the Harmonic Model or the Quasi-Harmonic Model~\cite{Pantazis:08c}, using the Least-Squares minimization method. However, the basis functions are both non-parametric and non-stationary:
\begin{equation}
\begin{bmatrix}
\mathbf{\hat{a}}\\
\mathbf{\hat{b}} \end{bmatrix} = (\mathbf{E_e}^H\mathbf{W}^H\mathbf{WE_e})^{-1}\mathbf{E_e}^H\mathbf{W}^H\mathbf{Ws}
\label{eq:LSeaQHM}
\end{equation}
where $\mathbf{\hat{a}}, \mathbf{\hat{b}}$ are the parameter vectors, $\mathbf{W}$ is the matrix containing the window values in the diagonal, $s$ is the input signal vector, the matrix $\mathbf{E_e}$ is defined as $\mathbf{E_e} = [\mathbf{E_{e0}}|\mathbf{E_{e1}}]$, and the submatrices $\mathbf{E_{ei}}$, $i=0,1$ have elements given by 
\begin{equation}
(E_{e0})_{n,k} = \hat{\alpha}_k(t_n)e^{j\hat{\phi}_k(t_n)}
\end{equation}
and
\begin{equation}
(E_{e1})_{n,k} = t_n\hat{\alpha}_k(t_n)e^{j\hat{\phi}_k(t_n)} = t_n(E_{e0})_{n,k}
\end{equation}
It can be observed that the basis functions are adapted to the local amplitude and frequency characteristics of the signal. Parameters $\hat{\alpha}_k(t)$ and $\hat{\phi}_k(t)$ are iteratively refined using $a_k$ and $b_k$, which when combined, they form a frequency correction term for each sinusoid, first introduced in~\cite{Pantazis:08c}:
\begin{equation}
\hat{\eta}_k = \frac{1}{2\pi}\frac{\Re\{a_k\}\Im\{b_k\} - \Im\{a_k\}\Re\{b_k\}}{|a_k|^2}
\end{equation}
where $\hat{\eta}_k$ is an estimate of the frequency mismatch between the actual frequency of the $k^{th}$ sinusoid and the estimated one. Applying the $\hat{\eta}_k$ on each frequency track, interpolating the instantaneous parameters over successive frames and restructuring the basis functions leads to more accurate model parameters $a_k$ and $b_k$. These form a new frequency mismatch correction, $\tilde{\eta}_k$. This way, the loop goes on until the instantaneous parameters yield a close representation of the underlying signal, according to a criterion based on the Signal-to-Reconstruction-Error Ratio (SRER), defined as:
\begin{equation}
SRER = 20\log_{10}\frac{std(x(t))}{std(x(t) - s(t))}
\label{eq:SRER}
\end{equation}
where $std(\cdot)$ is the standard deviation, $x(t)$ is the original signal, and $s(t)$ is the reconstructed signal. The adaptation algorithm can be found in~\cite{Kafentzis:12}.

\begin{figure}[htb!]
\centering
\subfloat[][Frequency adaptation of a sinusoid inside the analysis window.]{\includegraphics[width=1\linewidth]{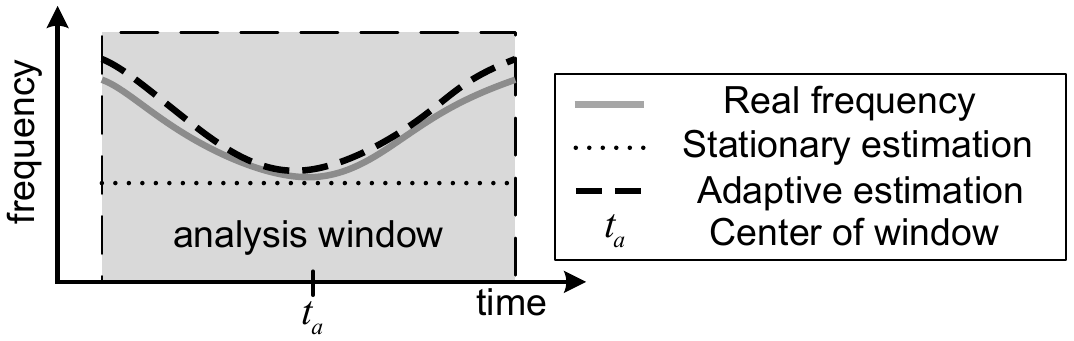}\label{fig:insidewinfr}}

\subfloat[][Amplitude adaptation of a sinusoid inside the analysis window.]{\includegraphics[width=1\linewidth]{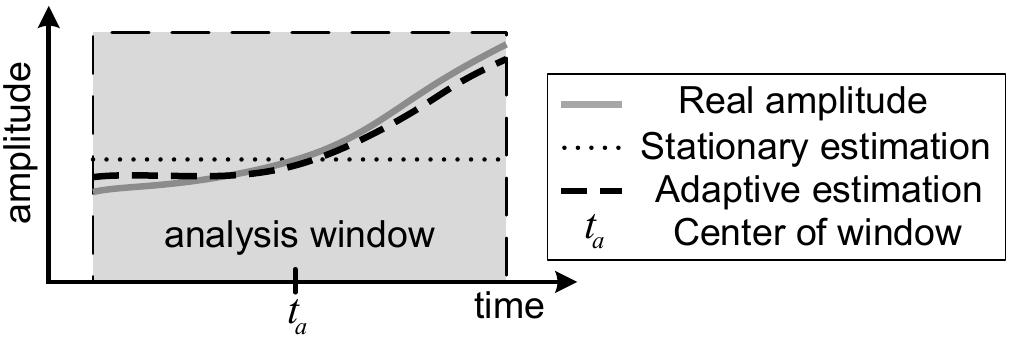}\label{fig:insidewinamp}}
\caption{{\it Inside the analysis window, the frequency (a) and amplitude (b) trajectories of a partial (solid grey line) is assumed to be constant for a stationary sinusoidal model (dotted line), while eaQHM (dashed line) iteratively adapts to the shape of the instantaneous component.}}
\end{figure}

In picturing the amplitude and frequency modelling of eaQHM within the analysis window, Figures~\ref{fig:insidewinfr} and~\ref{fig:insidewinamp} show how conventional (stationary) sinusoidal models like HM, SM, or QHM behave inside their analysis window. Their exponential basis functions are stationary in frequency, thus being inefficient on the representation of highly non-stationary frequency curves. The same argument applies for amplitude curves, although frequency estimation is far more important than amplitude estimation.
After applying the eaQHM for a number of adaptations, the instantaneous parameters are interpolated over successive frames and the overall signal is synthesized as in Eq.~(\ref{eq:SM}). For more insight on eaQHM, please refer to~\cite{Kafentzis:12, Kafentzis:aSM:14}.

\section{Experimental Results}
\label{eq:eval}
In this section, two separate experiments are presented. First, the models are evaluated on the estimation of the parameters of synthesized signals with variable analysis window size. This way, the features of each model are revealed. Second, an evaluation on real speech and audio data is presented, where the performance of each model is evaluated perceptually and numerically.

\subsection{Synthetic Signals}
The algorithms will be tested on synthetic signals in order to show the properties of each model, along with their corresponding advantages. The experimental set-up is the following: for the SM, a $2048$ FFT is used. Cubic interpolation for the phases and linear interpolation for the amplitudes is used. A hop size of $1$ ms is selected and the window is of Hamming type. For the eaQHM, the amplitudes are linearly interpolated while the frequencies are spline-interpolated. The phases are estimated as the integral of the instantaneous frequency as described in~\cite{Kafentzis:12}. Finally, the analysis window is of Hamming type.

\subsubsection{Mono-component signal}
\begin{figure}[!htb]
\centerline{\includegraphics[width=100mm]{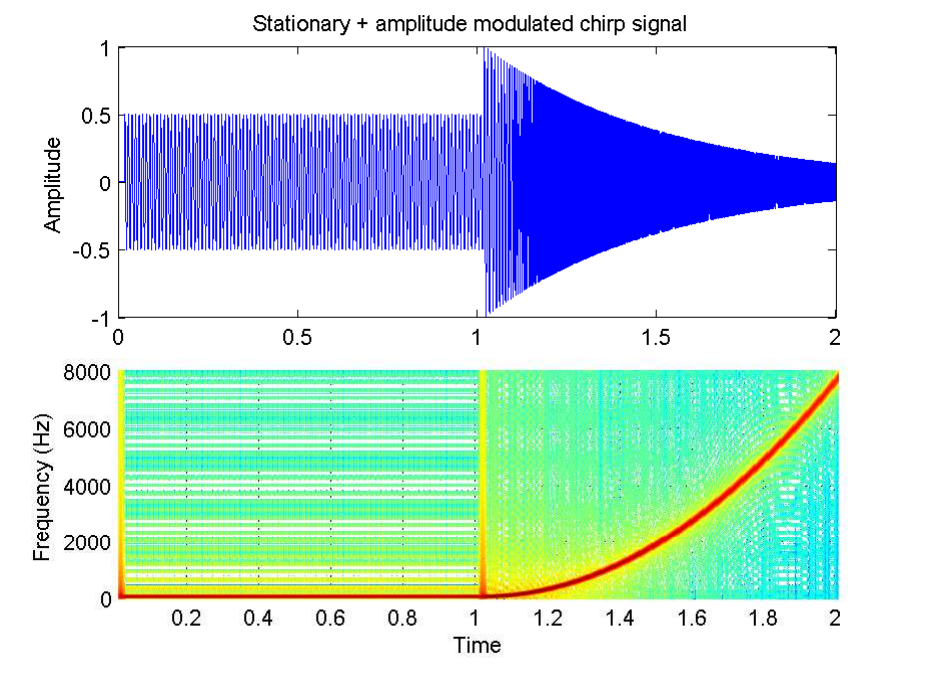}}
\caption{{\it Stationary sinusoid plus exponentially amplitude-modulated chirp signal. Upper panel: time domain. Lower panel: frequency domain.}} 
\label{fig:chirp}
\end{figure}

In Figure~\ref{fig:chirp}, a concatenated signal is presented. It is constructed by a stationary in amplitude and frequency sinusoid ($f_0 = 100$ Hz) with $1$ sec duration, and an exponentially amplitude-modulated chirp signal, starting from $100$ Hz up to $1000$ Hz, with its damping factor as $d=-2$ and its duration equal to $1$ sec as well. The sampling frequency is set to $f_s = 16$ kHz. So, the signal is simply one partial that is amplitude- and frequency- modulated, however its sharp transient part is suitable for comparing the performance of two models (eaQHM, EDSM) that both claim to handle sharp onsets well~\cite{Caetano:13a, Jensen:99, Kafentzis:12b}.

\begin{figure}[!htb]
\centerline{\includegraphics[width=95mm]{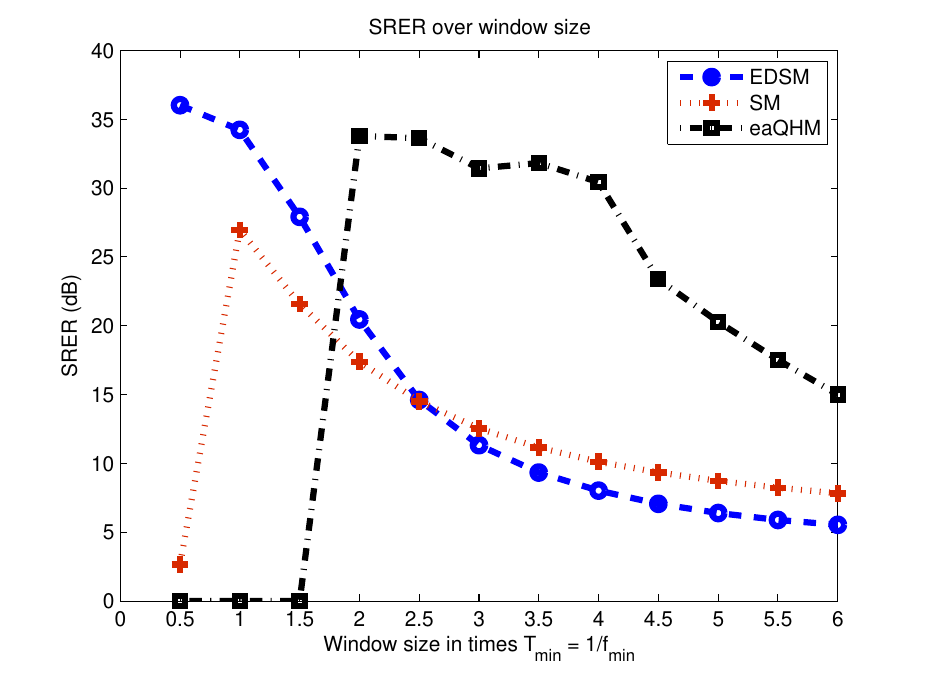}}
\caption{{\it Signal-to-Reconstruction-Error Ratio over analysis window size for single-component signal. Window size is a multiple of the minimum period of the signal, namely $T_{min} = 10$ ms.}} 
\label{fig:SRERwinsize}
\end{figure}

We will test the performance of the models using variable window size but seeking only one partial on the time-frequency domain. It should be mentioned that since LS is in the heart of eaQHM estimation scheme, the window size should be equal or greater than $\hat{T} = 2f_s/f_{min}$, otherwise the estimation suffers from conditioning problems. That poses a disadvantage compared to EDS and SM, where the window size can reach lower values. The minimum frequency in this signal is $100$ Hz, so the eaQHM estimation will start from the corresponding window size, which is $2T_{min} = 2f_s/f_{min}$ samples. The results are shown in Figure~\ref{fig:SRERwinsize}.

Please note that zero values for the eaQHM in the first three multiples of $T_{min}$ is due to ill-conditioning, so no result is obtained. However, when the window size is long enough, it can be seen that the eaQHM can outperform both sinusoidal models. This is due to the fact that the parameters are estimated on an set of \textit{adaptive} basis functions, which are formulated according to the amplitude and frequency changes \textit{inside} the analysis window. Furthermore, when the window size increases, the basis functions are averaged over time and the model adapts much slower. 
On the other hand, the EDSM manages to achieve high SRER values for window sizes where eaQHM parameter estimation is unstable. This is expected since the EDSM uses stationary sinusoids inside the analysis window, and the smaller the window size, the more valid is the assumption of stationarity inside it. Not surprisingly, the EDSM can approach the eaQHM in terms of SRER when more partials per window are allowed. In this case, EDS approximates the frequency modulation by a Bessel function-weighted cosine series~\cite{Carlson:09}. 
Finally, the SM behaves as expected; a small window fails to capture the stationary part at $100$ Hz, while performs well for the modulated chirp signal (except for the transient part), an optimal window maximizes the time-frequency trade-off, and a large window captures the stationary part well but smears out the energy in the transient part, and also the frequency-varying part is poorly modeled due to the effect of averaging.

\subsubsection{Multi-component signal}
\begin{figure}[!htb]
\centerline{\includegraphics[width=95mm]{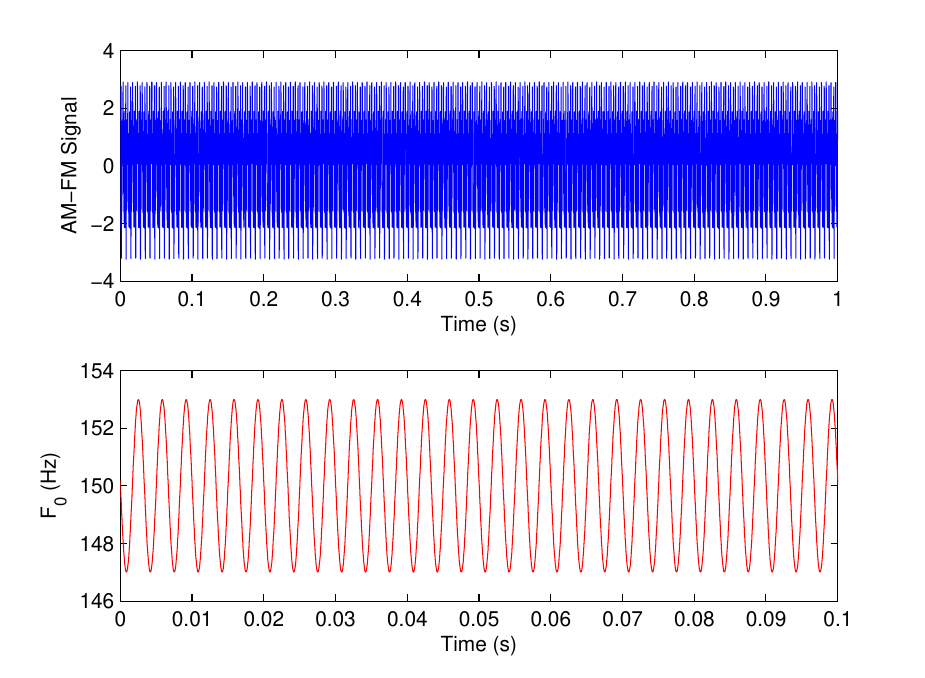}}
\caption{{\it Multicomponent AM-FM signal with sinusoidal frequency modulation. Upper panel: Signal in time domain, Lower panel: instantaneous frequency.}} 
\label{fig:AMFM}
\end{figure}

\begin{figure}[!htb]
\centerline{\includegraphics[width=95mm]{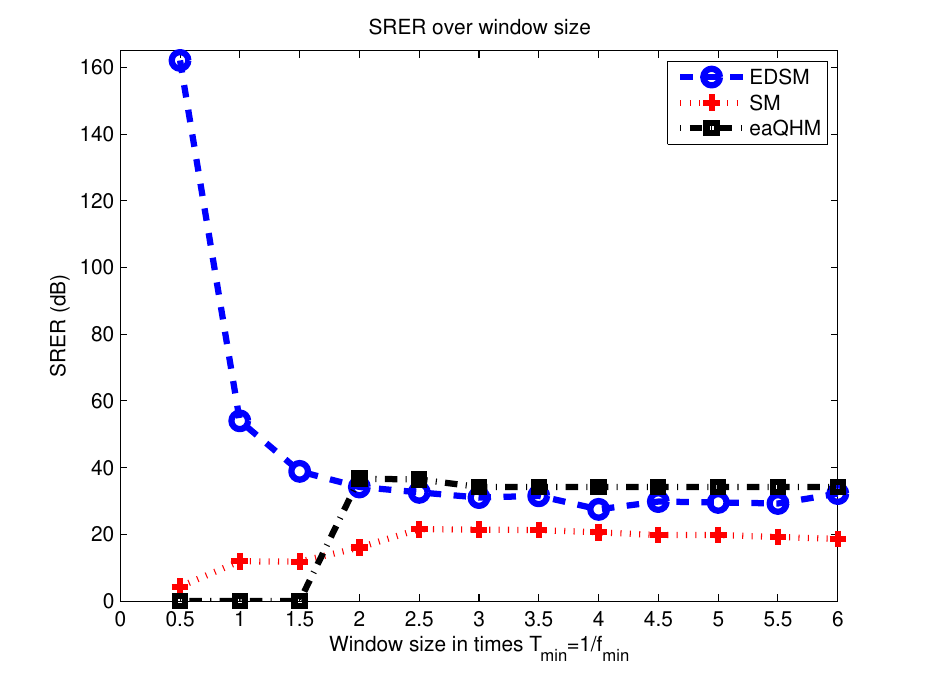}}
\caption{{\it Signal-to-Reconstruction-Error Ratio over analysis window size for multi-component signal. Window size is a multiple of the minimum period of the signal, namely $T_{min} = 6.68$ ms.}} 
\label{fig:SRERwinsizeAMFM}
\end{figure}

Let us try now a multicomponent AM-FM signal with $10$ partials, with sinusoidal frequency modulation, formulated as
\begin{equation}
x(t) = \sum_{k=1}^{10} A_k(t)\cos(\phi_k(t))
\label{eq:AMFMexample}
\end{equation}
with
\begin{equation}
A_k(t) = \frac{1}{2}+\frac{r}{k} , \: \: \: \phi_k(t) = 2\pi kf_0 t + k\rho \cos(2\pi f_c t)
\end{equation}
where $r$ is a random number in $[0,1]$, $f_0 = 150$ Hz, $f_c = 300$ Hz, and $\rho =0.01$. This signal is depicted in Figure~\ref{fig:AMFM}. The signal duration is $1$ sec the sampling frequency is $16$ kHz.
Figure~\ref{fig:SRERwinsizeAMFM} shows the evolution of SRER over time for the signal of Eq.~(\ref{eq:AMFMexample}). Again, the eaQHM is estimated only when there is no conditioning problem in parameter estimation, that is when the length of the analysis window is equal to or greater than $\hat{T} = 2T_{min} = 2f_s/f_{min}$. It can be seen that EDSM is able to reach an SRER of $165$ dB for a small window size of only $T_{min}/2 = 3.4$ ms), and decreasing when the window size increases. When the eaQHM is able to reliably estimate the signal parameters, then it outperforms EDSM, even for larger window sizes, by an average of $6.2$ dB. Finally, the SM behaves again as expected. Small windows yield larger mainlobes in frequency, and thus poor frequency discrimination, whereas larger windows give smaller mainlobes but as the window size increases, averaging effects on the spectrum result in decreasing reconstruction performance. As in the mono-component case, there is an window size where the time-frequency trade-off is optimal.

\subsection{Real Signals}
In the previous section, it was shown that the EDSM performs its best when the analysis window size is small. The eaQHM suffers from the LS conditioning when the analysis window is smaller than an optimal. Thus, in running speech and audio analysis, window sizes should vary according to the local pitch period. Based on this, a full-band analysis-synthesis system with pitch-adaptive analysis window based on eaQHM is designed in~\cite{Kafentzis:aSM:14}. In this section, the performance of systems based on eaQHM and EDSM are compared on real speech and audio data. The evaluation will be done in terms of SRER (Eq.~(\ref{eq:SRER})). Since a complete evaluation regarding the number of partials, the hop size, and the window size is out of the scope of this work, the parameters will be fixed in an nearly optimal way.

For the evaluation, a small database of ten audio signals were used, varying from singing male and female voice to electric guitar solos and musical instruments with changing pitch over time. All sounds were sampled at $16000$ Hz, with $16$ bits of storage accuracy. Since our primary concern in this work is modelling accuracy, the experimental set-up is the following: the analysis window is $30$ ms and $3$ local pitch periods for SM and eaQHM respectively, whereas the hop size is $1$ ms for both models. The number of partials is set to cover the full-band for eaQHM (according to the estimated $f_0$) in each window, whereas for the SM, at most $100$ partials where sought in the peak-picking process of a $2048$-point FFT magnitude spectrum. The eaQHM uses a Blackman window for the estimation of an initial harmonic template, and a Hamming window for the adaptation part, while the SM uses a Hann window. At most $10$ adaptations were allowed for the eaQHM. Instantaneous amplitudes are linearly interpolated in both models. Instantaneous frequencies are linearly interpolated in the SM whereas spline interpolation is used in the eaQHM. Cubic interpolation~\cite{McAulay:86} and frequency integration~\cite{Pantazis:11} are applied in the SM and eaQHM, respectively. For the EDSM, the window size is rectangular, fixed, and equal to $0.75$ times the average pitch period, as it is estimated by the SWIPE pitch estimator~\cite{Camacho:08}. The hop size is equal to the window length, thus there is no overlap between windows. The number of partials per frame depend on the local fundamental frequency, $f_0^l$, and is set to span the entire spectrum. Hence, the number of partials per frame is equal to $f_s/(2f_0^l)$.

It is important to note that the number of synthesis parameters is equal for the SM and eaQHM (amplitudes $A_k$, frequencies $f_k$, and phases $\phi_k$), since both of them result in the same synthesis equation~(\ref{eq:SM}), whereas for the EDS the number of synthesis parameters is larger by one (amplitudes $|a_k|$, frequencies $\omega_k$, phases $\phi_k$, damping factors $\delta_k$).

Table~\ref{tab:SRER1} shows the SRER values obtained for the eaQHM, the EDS, and the baseline SM for the aforementioned database. 
\begin{table}[!htb]
\begin{center}
\begin{tabular}{|c|c|c|c|c|c|}
\cline{1-6}
\multicolumn{6}{|c|}{SRER Performance} \\ \hline
Model & Bass 1 & Bass 2 & Soprano 1 & Soprano 2 & Violin\\ \hline
$SM$ & $13.8$ & $13.6$ & $17.6$ & $18.1$ & $17.8$\\ \hline
$EDS$ & $34.4$ & $34.4$ & $35.5$ & $35.6$ & $28.9$\\ \hline
$eaQHM$ & $ 31.3$ & $33.2$ & $34.6$ & $35.2$ & $32.7$\\ \hline\hline
Model & Guitar 1 & Guitar 2 & Vocals 1 & Vocals 2 & Harp\\ \hline
$SM$ & $12.2$ & $11.9$ & $18.9$ & $19.5$ & $15.3$\\ \hline
$EDS$ & $18.6$ & $15.7$ & $25.7$ & $27.8$ & $14.2$\\ \hline
$eaQHM$ & $27.8$ & $28.8$ & $28.2$ & $29.9$ & $20.2$\\ \hline
\end{tabular}
\end{center}
\caption{{\it Signal to Reconstruction Error Ratio values (dB) for all models on a database of $10$ audio signals.} }
\label{tab:SRER1}
\end{table}
From Table~\ref{tab:SRER1}, we notice that the eaQHM and the EDSM perform similarly and much higher than the SM for quasi-harmonic signals, such as male/female singing voices and violin solo. However, for signals with highly non-stationary content, such as an electric guitar solo, EDSM needs more partials or smaller window size in order to model the underlying signal characteristics. On the contrary, the eaQHM is able to adapt its parameters inside the analysis window, thus providing more accurate parameter estimation and higher reconstruction quality. 


\section{Discussion and Perspectives}
\label{eq:disc}
The results of the previous section clearly show that the eaQHM can provide high reconstruction quality for a variety of sounds that exhibit high non-stationarity over time. Although some of the audio signals evaluated in this work have no quasi-harmonic structure at all, adaptation allows the convergence of the harmonically-initialized model into the frequency content of the analyzed signal. Despite the lack of a noise component, informal listening tests showed that the perceptual similarity between the original and the reconstructed signal is very high. However, high quality comes with the cost of increased time complexity. On an Intel Core i$7$ CPU, with $6$ GB RAM, the standard SM models the audio signals in less than $5$ seconds on average, whereas the eaQHM averages about $3.5$ minutes, with a mean number of $4.2$ adaptations, for a $16$ kHz sampled, $16$ bit audio signal. EDSM is in between but rather fast, with a mean number of $12$ seconds per sound file. It is of interest to decrease the runtime of the eaQHM by applying faster methods for parameter estimation, such as the FFT. That would make the model more attractive for near-real-time applications. Adaptation based on a different estimation scheme than Least-Squares is under investigation.

\section{Conclusions}
In this paper, we compared three sinusoidal model variants, with different parameter estimation techniques. The advantages and the disadvantages of each one were demonstrated via synthetic and real signals of different content. Conclusively, the subspace method of EDS is very powerful in parameter estimation, despite of using stationary sinusoids inside the analysis window. The EDS has been successfully used in analysis of monophonic and polyphonic audio. On the other hand, the adaptive Sinusoidal models use a less powerful estimation method, the well-known LS minimization, but the ability to change the characteristics of their basis functions in the LS estimation provides a powerful iterative mechanism to obtain high accuracy instantaneous parameters. However, LS solutions suffers from ill-conditioning when two sinusoids have frequencies which are very close to each other, making the method inappropriate for polyphonic audio analysis. Moreover, the eaQHM has the same number of analysis parameters with the EDS in a single frame, namely four, but synthesizes a signal segment with three, instead of four, parameters. One possible future research direction is to combine the strong points of each method into a new paradigm, able to accurately estimate the parameters and in parallel, to adapt in any signal content.

\bibliographystyle{IEEEbib}
\bibliography{biblio} 

\end{document}